\documentstyle[12pt]{article}

\catcode`\@=11
\long\def\@makefntext#1{
\protect\noindent \hbox to 3.2pt {\hskip-.9pt

$^{{\ninerm\@thefnmark}}$\hfil}#1\hfill}        

 \def\@makefnmark{\hbox to 0pt{$^{\@thefnmark}$\hss}}  

\def\ps@myheadings{\let\@mkboth\@gobbletwo
\def\@oddhead{\hbox{}
\rightmark\hfil\ninerm\thepage}

\def\@oddfoot{}\def\@evenhead{\ninerm\thepage\hfil
\leftmark\hbox{}}\def\@evenfoot{}
\def\sectionmark##1{}\def\subsectionmark##1{}}

\newcounter{sectionc}\newcounter{subsectionc}\newcounter{subsubsectionc}
\renewcommand{\section}[1] {\vspace{0.6cm}\addtocounter{sectionc}{1}

\setcounter{subsectionc}{0}\setcounter{subsubsectionc}{0}\noindent

	{\bf\thesectionc. #1}\par\vspace{0.4cm}}
\renewcommand{\subsection}[1]
{\vspace{0.6cm}\addtocounter{subsectionc}{1}

	\setcounter{subsubsectionc}{0}\noindent

	{\it\thesectionc.\thesubsectionc. #1}\par\vspace{0.4cm}}
\renewcommand{\subsubsection}[1]
{\vspace{0.6cm}\addtocounter{subsubsectionc}{1}
	\noindent
{\rm\thesectionc.\thesubsectionc.\thesubsubsectionc.

	#1}\par\vspace{0.4cm}}

\newcounter{appendixc}
\newcounter{subappendixc}[appendixc]
\newcounter{subsubappendixc}[subappendixc]

\renewcommand{\appendix}[1] {\vspace{0.6cm}
	\refstepcounter{appendixc}
	\setcounter{figure}{0}
	\setcounter{table}{0}
	\setcounter{equation}{0}
	\renewcommand{\thefigure}{\Alph{appendixc}.\arabic{figure}}
	\renewcommand{\thetable}{\Alph{appendixc}.\arabic{table}}
	\renewcommand{\theappendixc}{\Alph{appendixc}}

\renewcommand{\theequation}{\Alph{appendixc}.\arabic{equation}}
    \noindent{\bf Appendix \theappendixc #1}\par\vspace{0.4cm}}

\def\abstracts#1{{
  \centering{\begin{minipage}{30pc}\tenrm\baselineskip=12pt\noindent
  \centerline{\tenrm ABSTRACT}\vspace{0.3cm}
  \parindent=0pt #1
  \end{minipage}}\par}}

\renewenvironment{thebibliography}[1]
	{\begin{list}{\arabic{enumi}.}
	{\usecounter{enumi}\setlength{\parsep}{0pt}
\setlength{\leftmargin 1.25cm}{\rightmargin 0pt}
	 \setlength{\itemsep}{0pt} \settowidth
	{\labelwidth}{#1.}\sloppy}}{\end{list}}

\newcounter{itemlistc}
\newcounter{romanlistc}
\newcounter{alphlistc}
\newcounter{arabiclistc}

\newcommand{\fcaption}[1]{
	\refstepcounter{figure}
	\setbox\@tempboxa = \hbox{\tenrm Fig.~\thefigure. #1}
	\ifdim \wd\@tempboxa > 6in
	   {\begin{center}
	\parbox{6in}{\tenrm\baselineskip=12pt Fig.~\thefigure. #1}
	    \end{center}}
	\else
	     {\begin{center}
	     {\tenrm Fig.~\thefigure. #1}
	      \end{center}}
	\fi}

\newcommand{\tcaption}[1]{
	\refstepcounter{table}
	\setbox\@tempboxa = \hbox{\tenrm Table~\thetable. #1}
	\ifdim \wd\@tempboxa > 6in
	   {\begin{center}
	\parbox{6in}{\tenrm\baselineskip=12pt Table~\thetable. #1}
	    \end{center}}
	\else
	     {\begin{center}
	     {\tenrm Table~\thetable. #1}
	      \end{center}}
	\fi}

\def\@citex[#1]#2{\if@filesw\immediate\write\@auxout
	{\string\citation{#2}}\fi
\def\@citea{}\@cite{\@for\@citeb:=#2\do
	{\@citea\def\@citea{,}\@ifundefined
	{b@\@citeb}{{\bf ?}\@warning
	{Citation `\@citeb' on page \thepage \space undefined}}
	{\csname b@\@citeb\endcsname}}}{#1}}

\newif\if@cghi
\def\cite{\@cghitrue\@ifnextchar [{\@tempswatrue
	\@citex}{\@tempswafalse\@citex[]}}
\def\citelow{\@cghifalse\@ifnextchar [{\@tempswatrue
	\@citex}{\@tempswafalse\@citex[]}}
\def\@cite#1#2{{$\null^{#1}$\if@tempswa\typeout
	{IJCGA warning: optional citation argument

	ignored: `#2'} \fi}}

\def\fnt#1#2{\footnotetext{\kern-.3em
	{$^{\mbox{\sevenrm #1}}$}{#2}}}

\font\twelvebf=cmbx10 scaled\magstep 1
 1
 1

\font\tenbf=cmbx10
\font\tenrm=cmr10
\font\tenit=cmti10

\font\ninerm=cmr9

\newcommand{\br}{{\bf r}}
\newcommand{\anu}{{|\nu|}}

\renewcommand{\u}{\Psi}

\newcommand{\be}{\begin{equation}}
\newcommand{\ee}{\end{equation}}
\newcommand{\bea}{\begin{eqnarray}}
\newcommand{\eea}{\end{eqnarray}}

\newcommand{\kk}{\rho}

\begin{document}

\centerline{\tenbf FIELD THEORETICAL AND QUANTUM MECHANICAL
DESCRIPTIONS}
\baselineskip=16pt
\centerline{\tenbf OF COLLIDING AND NON-COLLIDING ANYONS}
\vspace{0.8cm}
\centerline{\tenrm Giovanni AMELINO-CAMELIA}
\baselineskip=13pt
\centerline{\tenit Center for Theoretical Physics, Building 6,}
\baselineskip=12pt
\centerline{\tenit Laboratory for Nuclear Science,
and Department of Physics}
\baselineskip=12pt
\centerline{\tenit Massachusetts Institute of Technology}
\baselineskip=12pt
\centerline{\tenit Cambridge, Massachusetts 02139, USA}
\vspace{0.9cm}
\abstracts{I discuss two techniques
that can be used in the investigation of the properties
of ``colliding" and ``non-colliding" anyons.}

\vfil
\rm\baselineskip=14pt
\section{Introduction}
Several definitions of anyons\cite{wil} have been given in the
literature (for a recent analysis
of the different possible definitions
of these planar particles, see Ref.[2]).
In this paper
I discuss anyons in the ``boson gauge"\cite{wil,gac},
and therefore they are described as bosons
interacting
through the mediation of an
abelian Chern-Simons gauge field,
{\it i.e.} the ``free anyon Hamiltonian" is
(for anyons of unit mass)
\bea
&H = {1 \over 2} \sum_n \biggl( {\bf p}_n - \nu {\bf a}_n \biggr)^2
{}~,& \label{hany}\\
&a_n^k \equiv - \epsilon^{kj} \sum_{m (\ne n)} {r_n^j - r_m^j \over
|{\bf r}_n - {\bf r}_m|^2 } ~,~~~ 0 < \nu < 1
{}~,& \nonumber
\eea
where ${\bf r_n} \equiv (r_n^1, r_n^2)$
is the position vector of the n-th anyon, and
$\nu$ is the ``statistical parameter", which characterizes
the anyon statistics\cite{wil,gac}.

In the plane also fermions can be described as bosons interacting as
in (\ref{hany}) with the particular choice of the
statistical parameter $\nu = 1$.
Ordinary planar bosons obviously correspond to
(\ref{hany}) with $\nu = 0$.
The fact that the (non-integer) values of
the statistical parameter $\nu$
that correspond to anyons interpolate
between the bosonic ($\nu = 0$)
and the fermionic ($\nu = 1$) limit plays an important role
in anyon physics, and I exploit it in the following.

Whereas for bosons (fermions) it is well understood
that the wave functions
do not have to (have to) vanish at the ``points of overlap", which are
the points of configuration space where some of the particle
positions coincide,
the situation is more complex in the case of anyons\cite{bour}.
For conceptual simplicity,
in a large majority of studies only ``non-colliding anyons"
({\it i.e.} anyons whose wave function vanish at the points of overlap)
have been considered;
however, in this paper anyonic wave functions are only required to be
consistent with the physical conditions that they be
square integrable and diverge at most at a finite number of points,
and the Hamiltonian be self-adjoint\footnote{In the case of
free bosons and free
fermions these physical conditions are sufficient to
determine\cite{bour} the behavior
of the wave functions at the points of
overlap.}.
In particular, as discussed in
Refs.[5,6],
these conditions
imply that the wave functions\footnote{Note that
one can easily see\cite{manu2} that for non-s-wave functions
square integrability is only consistent with the $\psi(0)=0$
boundary condition; therefore, these generalized boundary conditions
only affect
the s-wave part of calculations.}
describing the relative
motion of two anyons must satisfy the following boundary
condition
at the point
of overlap ($r = 0$)
\begin{equation}
\left[r^\anu\psi(\br) - w R^{2\anu}{d\left(r^\anu \psi (\br)\right)
\over d (r^{2\anu})}\right]_{r=0}=0 ~,
\label{bc1}
\end{equation}
which can be equivalently expressed as the following requirement on
the form of $\psi$ for ${r} \sim 0$
\begin{equation}
\psi(\br)\rightarrow a(r^\anu + w R^{2\anu}r^{-\anu})\ \ \
{\rm for} \ r \sim 0 ~.
\label{bc2}
\end{equation}
$R$, the ``self-adjoint extension
scale"\footnote{As clarified in
Refs.[5,6],
this discussion of the boundary conditions
for anyons is an application
of the method of self-adjoint
extension of the Schr\"odinger Hamiltonian.},
is a reference scale with dimensions of a length,
$w$, the ``self-adjoint extension parameter",
is a dimensionless real parameter
which characterizes
the type of boundary condition, and $a$ is a constant.
Note at the ``critical points" $w \! = \! 0$,
which corresponds to the
conventional non-colliding anyons, and $w \! \sim \! \infty$ the
theory becomes scale invariant ({\it i.e.} it is
independent of the scale $R$).

I am now ready to give more rigorously the definition of anyons
adopted in this paper; they are
particles described by wave functions
satisfying boundary conditions of the type (\ref{bc1}),
whose free evolution is
governed by $H$ given in (\ref{hany}).
(Obviously in presence of interactions described by a potential $V$
the evolution is governed by the Hamiltonian $H+V$.)
Given a self-adjoint extension scale $R$
the anyons defined in this way
are characterized by two numbers, the statistical parameter $\nu$
and the self-adjoint extension parameter $w$.
In the following
I discuss a field theoretical and a quantum mechanical
method of investigation of the properties of these particles.

\section{Non-Relativistic Field Theory}
In this section,
I discuss a field theoretical
method of investigation of anyons.
Let me start by considering the
Lagrange density
\begin{eqnarray}
 {\cal L}= {1 \over 4 \pi \kk}\epsilon^{\alpha\beta\gamma}
A_\alpha\partial_\beta A_\gamma
 +i\phi^\dagger D_t \phi -
 {1\over 2}({\bf D}\phi)^\dagger\cdot {\bf D}\phi ~,
\label{a-lag}
\end{eqnarray}
where $\phi$ is a complex
bosonic field,
$({\bf A},A_0)$ is an auxiliary\cite{jackpi}
abelian Chern-Simons gauge field,
and $D_t \! \equiv \! \partial_t \! + \! i A_0$ and
$ {\bf D} \! \equiv \! \nabla \! - \! i {\bf A}$
are the covariant derivatives.

In Refs.[8,9]
it was shown that
renormalizability requires that
a contact term $-\pi g_b (\phi^\dagger\phi)^2$ be
added to ${\cal L}$.
The renormalized s-wave two-particle
scattering amplitude was calculated
to two-loop order
in Ref.[6];
including the appropriate kinematic factor it
can be written as
\begin{eqnarray}
&A_{s,2-{\rm loop}} \! = \! -{\sqrt{{2\pi \over p}}} \{
g_r \!-\! {i \pi \over 2} \kk^2 \!-\! {\pi^2 \over 6} g_r \kk^2
\!+\! (g_r^2 \!-\! \kk^2)
\! \left(\ln {p\over {\mu}}
\!-\! {i\pi\over 2} \right)
\!+\! g_r(g_r^2 \!-\! \kk^2)
\! \left(\ln {p\over {\mu}} \!-\! {i\pi\over 2} \right)^2 \} , \,~&
\label{a-peramp2}
\end{eqnarray}
where ${\bf p}$ is the relative momentum,
$\epsilon$ is the usual cut-off used in dimensional regularization,
$\mu$ is the renormalization scale,
and $g_r$ is the
two-loop renormalized contact coupling
which is defined in terms of the
bare contact coupling $g_b$ by
the relation ($\gamma$ denotes the Euler constant)
\begin{eqnarray}
&g_r = g_b - (g_b^2-\kk^2)({1\over 2\epsilon}
- {\gamma -\ln 4\pi\over 2})
+g_b(g_b^2-\kk^2)({1\over 2\epsilon}
- {\gamma -\ln 4\pi\over 2})^2 ~,&
\label{a-reno}
\end{eqnarray}
Note that only at the
critical values $g_r=\pm \kk$ of the renormalized
contact coupling the
scale invariance of the classical theory is preserved at the
quantum level.

In order to show how to use the results
of this field theory  in the study of anyons, let me compare
Eq.(\ref{a-peramp2}) to the exact s-wave scattering amplitude
of anyons.
This scattering amplitude
can be evaluated exactly by using
a rather straightforward generalization of the analysis given
in Ref.[10],
which concerned the special case $w=0$;
one finds\cite{gacbak} that
\begin{eqnarray}
&A_s(p) = - i {\sqrt{{2 \over \pi p}}}
(e^{i\pi\anu} -1){1-{1 \over w}
\left({2\over pR}\right)^{2\anu}{\Gamma(1+\anu)\over
\Gamma(1-\anu)}\over 1+ {1 \over w}
e^{i\pi\anu}\left({2\over pR}\right)^{2\anu}{\Gamma(1+\anu)\over
\Gamma(1-\anu)}} ~~~~~~~~~~ ~~~~~~~~~~~~&
\nonumber\\
&= -{\sqrt{{2\pi \over p}}} \{ \anu {1-w\over 1+w}
-{i \pi \over 2}\nu^2
- \nu^2{4w\over (1+w)^2}\left(\ln {pR\over 2}+\gamma -
{i\pi\over 2}\right)   \nonumber\\
&~~~~~~ ~~~~~~~~~~~~ ~~~~\,
- {\pi^2 \over 6}\anu^3 {1-w\over 1+w}
\!-\! \anu^3{4(1-w)w\over (1+w)^3}\left(\ln {pR\over 2}\!+\!
\gamma\! -\!{i\pi\over 2}\right)^2\!\! +\! O(\nu^4)
\} ~.&
\label{a-per1}
\end{eqnarray}

If one uses the relations
\begin{eqnarray}
\kk = \nu ~,~~~~ g_r= |\nu| {1-w \over 1+w} ~,~
\label{a-rel1}
\end{eqnarray}
\begin{eqnarray}
\mu = {2\over Re^{\gamma}}
\label{a-rel2}
\end{eqnarray}
the two-loop field theoretical
result (\ref{a-peramp2}) reproduces exactly the $O(\nu^3)$
approximation
of the exact result (\ref{a-per1}).
This indicates that the field theory discussed in this section
describes anyons of statistical
parameter $\kk$ and self-adjoint extension parameter
\space\space\space\space\space\space\space\space\space\space\space
$(|\kk| - g_r)/(|\kk| + g_r)$ at a self-adjoint extension
scale $2 \mu^{-1} e^{- \gamma}$.

Further insight into the correspondence
between the anyon quantities $w$,$R$ and the field theoretical
quantities $g_r$,$\mu$
can be gained from the following observations.
First, notice that, using
the renormalization-group
equation which states that the
scattering amplitude obtained in field theory is independent of the
choice of the renormalization scale $\mu$, one can derive
the following beta function for the
coupling $g_r$
\begin{eqnarray}
\beta(g_r)\equiv {d g_r\over d \ln \mu} = g_r^2 -\kk^2 ~.
\label{a-beta}
\end{eqnarray}
Eq.(\ref{a-beta}), which indicates that $g_r$ and $\mu$
are not physically
independent, can be integrated to give the relation
\begin{eqnarray}
 {|\kk| +g_r(\mu_1) \over |\kk| -g_r(\mu_1)}\,\mu_1^{2 |\kk|}
={|\kk| +g_r(\mu_2) \over |\kk| -g_r(\mu_2)}\,\mu_2^{2 |\kk|} ~.
\label{a-beta1}
\end{eqnarray}
Similarly in the exact result (\ref{a-per1})
$R$ is only a reference scale, and obviously
physics must be independent of the choice of $R$. Indeed, all physical
quantities (see, for example, Eqs.(\ref{bc1}) and (\ref{a-per1}))
depend on $w$ and $R$ only through the product $w R^{2\anu}$,
and the independence of physics on the choice of $R$ is in
the fact that if $R$ is changed from a value $R_1$ to a value $R_2$
this must be accompanied by a corresponding change of $w$
as described by the
relation
\begin{eqnarray}
w(R_1) \, R_1^{2\anu} = w(R_2) \, R_2^{2\anu} ~.
\label{a-beta2}
\end{eqnarray}
Clearly, the Eqs.(\ref{a-beta1})
and (\ref{a-beta2}) are perfectly consistent
with the relations (\ref{a-rel1}) and (\ref{a-rel2}).

\section{Quantum Mechanics}
The ideas discussed in the preceding section are also useful
in quantum mechanics,
leading to a quantum mechanical approach to the study of
anyons.
For simplicity in this section I discuss this approach only
in the case of the conventional non-colliding anyons ($w=0$),
which allows a scale invariant analysis.
However, one can straightforwardly
verify that, like in the field theory case, the approach can be
generalized to the study of ``colliding anyons" ($w \ne 0$).

Let me start by considering the Hamiltonian
\be
H_2 = H_{2}^{free} + r^2 =
- {1 \over r} \partial_{r} (r \partial_{r})
- {1 \over r^2}  \partial_{\phi}^{2}
+ r^2
-{ 2 i \kk \over  r^2}\partial_{\phi} + { \kk^2 \over r^2} ~.
\label{eqbe}
\ee
$H_{2}^{free}$ is the two-body relative motion Hamiltonian
of the quantum mechanical formu-
lation\cite{jackpi} of the
non-relativistic field theory
with Lagrangian ${\cal L}$ (see Eq.(\ref{a-lag}))
discussed in the preceding
section.
The $r^2$ term describes an harmonic interaction and discretizes the
spectrum.

In the perturbative calculations about $\kk=0$,
one runs into
an inconsistency because the matrix elements of
${\kk^2 \over r^2}$ between 0-th
order in $\kk$  s-wave functions
({\it i.e.} s-wave functions of bosons in an harmonic potential)
are logarithmically divergent; for example, to second order in
$\kk$, one of the contributions to the energy
of the state with $n=2$ and $l=0$
is given by ($L_{n}^{x}$ are Laguerre polynomials)
\bea
<\Psi^{(0)}_{2,0}|
{\kk^2 \over r^2} |\Psi^{(0)}_{2,0}> \!\!\!&=&\!\!\!
\int^{\infty}_{0} \int^{2\pi}_{0} r~dr~d\phi
{}~{e^{ - {r^2  \over 2}} \over \pi^{1 \over 2}}~L_2^0(r^2)~
{\kk^2 \over r^2}
{}~{e^{ - {r^2  \over 2}}
\over \pi^{1 \over 2}}~L_2^0(r^2)~     \nonumber\\
\!\!\!&=&\!\!\! 2 \kk^2
\int^{\infty}_{0} {\exp( -r^2 ) \over r}~dr~
\sim \infty ~.
\label{eqbf}
\eea

One can easily see$^{11-15}$
that these divergences are closely
related to the ultraviolet divergences encountered in the
field theoretical
calculations based on the Lagrangian of Eq.(\ref{a-lag}),
and it is therefore not surprising that they also can be cured by
introducing a contact interaction$^{11-15}$.
The renormalizable Hamiltonian is
\be
H^{ren}_2 \equiv
- {1 \over r} \partial_{r} (r \partial_{r})
- {1 \over r^2}  \partial_{\phi}^{2} + r^2
-{ 2i\kk \over  r^2}\partial_{\phi}
 + 2 \pi g_b \delta^{(2)}({\bf r})
+ { \kk^2 \over r^2}
=H_2 + 2 \pi g_b \delta^{(2)}({\bf r})
{}~,
\label{hamdel}
\ee
where $2 \pi g_b \delta^{(2)}({\bf r})$ is the
two-body quantum mechanical
counterpart of the $-\pi g_b (\phi^\dagger\phi)^2$ contact term
of the field theory discussed in the preceding section.

Unlike $H_2$, $H_2^{ren}$ is suitable for perturbation theory;
in fact,
with an appropriate choice of the contact coupling
(which corresponds to the
identifications (\ref{a-rel1}) and (\ref{a-rel2})),
the added $\delta$-function potential leads to
divergencies which cancel those introduced by the $\kk^2 / r^2$
term.

Let me look at some calculations
for the state with $n=2$ and $l=0$,
which will also illustrate this mechanism
of cancellation of divergencies.
The first order energy is given by
\bea
E^{(1)}_{2,0}
\!\!&=&\!\! <\Psi^{(0)}_{2,0}|
-{ 2i \kk \over  r^2}\partial_{\phi}
 + 2 \pi g_b \delta^{(2)}({\bf r})
|\Psi^{(0)}_{2,0}> \nonumber\\
\!\!&=&\!\! <\Psi^{(0)}_{2,0}| 2 \pi g_b \delta^{(2)}({\bf r})
|\Psi^{(0)}_{2,0}>
= 2 g_b
{}~.
\label{eonecoppia}
\eea

Concerning the first order eigenfunction one easily finds
\bea
|\Psi^{(1)}_{2,0}> \!\!& = &\!\!
\sum_{m,l \ne 2,0}
{<\Psi^{(0)}_{m,l}| ~
-{ 2i\kk \over  r^2}\partial_{\phi}
 + 2 \pi g_b \delta^{(2)}({\bf r}) ~ |\Psi^{(0)}_{2,0}>
\over E^{(0)}_{2,0}-E^{(0)}_{m,l} } |\Psi^{(0)}_{m,l}>
\nonumber\\
\!\!&= &\!\! \sum_{m \ne 2} {<\Psi^{(0)}_{m,0}| ~
2 \pi g_b \delta^{(2)}({\bf r}) ~
|\Psi^{(0)}_{2,0}> \over E^{(0)}_{2,0}-E^{(0)}_{m,0} }
|\Psi^{(0)}_{m,0}>
\nonumber\\
\!\!& = &\!\! - {g_b \over 2\sqrt{\pi }} \sum_{m \ne 2}
{L_{m}^{0}(r^2) \over m-2} e^{-{r^2 \over 2}}
\nonumber\\
\!\!& = &\!\! {g_b \over \sqrt{\pi }}  e^{-{r^2 \over 2}}
\biggl[ {3 \over 2} - r^2 + {1 \over 4} (2\gamma - 3 + 4 ln(r))
L_{2}^{0}(r^2) \biggr]
{}~.
\label{psioneto}
\eea

The second order energy is given by
\bea
E^{(2)}_{2,0}
\!\!\!&=&\!\!\! E^{(2,a)}_{2,0} ~ + ~
E^{(2,b)}_{2,0} ~, \label{etwocoppia}
\eea
\bea
E^{(2,a)}_{2,0} \!\!\! &=&
\!\!\! <\Psi^{(0)}_{2,0}| ~ {\kk^2 \over r^2} ~
|\Psi^{(0)}_{2,0}>  =
\kk^2 \int^{\infty}_{0} \int^{2 \pi}_{0} dr~ d\phi ~
{\exp( -r^2 ) \over \pi r}~ [L_{2}^{0}(r^2)]^2
\nonumber\\
&=& \!\!\!  \lim_{\epsilon \rightarrow 0}
\int^{\infty}_{\epsilon}  dr \, \kk^2 \,
{\exp( -r^2 ) \over r} \,
[L_{2}^{0}(r^2)]^2 \, = \,
\kk^2 \, \lim_{\epsilon \rightarrow 0}
\biggl[-2 \ln(\epsilon) -\gamma -{3 \over 2}\biggr]
\, ,
\label{etworega}
\eea
\bea
E^{(2,b)}_{2,0} \!\!\! &=& \!\!\! <\Psi^{(0)}_{2,0}| ~
-{ 2i\kk \over r^2}\partial_{\phi}
 + 2 \pi g_b \delta^{(2)}({\bf r})
{}~ |\Psi^{(1)}_{2,0}> ~
= ~ <\Psi^{(0)}_{2,0}| ~ 2 \pi g_b \delta^{(2)}({\bf r}) ~
|\Psi^{(1)}_{2,0}>
\nonumber\\
\!\!\! &=& \!\!\! 2 g_b^2
\int^{\infty}_{-\infty} \int^{\infty}_{-\infty}  dr_x \, dr_y \,
\delta^{(2)}({\bf r}) \, e^{-r^2} \, L_{2}^{0}(r^2) \,
\biggl[ {3 \over 2} - r^2 + {1 \over 4} (2\gamma - 3 + 4 ln(r))
L_{2}^{0}(r^2) \biggr]
\nonumber\\
\!\!\! &=& \!\!\! g_b^2 \, \lim_{\epsilon \rightarrow 0}
\biggl[2 ln(\epsilon) + {3 \over 2} + \gamma \biggr]
{}~.
\label{etworegb}
\eea

\noindent
Note that I introduced a cut-off $\epsilon$
(which must be ultimately removed by
taking the limit $\epsilon\rightarrow 0$)
in order to see the cancellation of infinities and
evaluate the left-over
finite result.
In general a similar cut-off must be introduced in all the divergent
matrix elements
of $r^{-2}$ and $\delta^{(2)}({\bf r})$
by using
\be
\int^{\infty}_{-\infty} \int^{\infty}_{-\infty}  dr_x ~ dr_y
{1 \over r^2}~f(r_x,r_y) =
\lim_{\epsilon \rightarrow 0}
\int^{\infty}_{\epsilon} \int^{2 \pi}_{0} r ~ dr ~ d\phi ~
{1 \over r^2}~f(r \cos \phi,r \sin \phi)
{}~,
\label{rega}
\ee
\be
\int^{\infty}_{-\infty} \int^{\infty}_{-\infty}  dr_x ~ dr_y
\delta^{(2)}({\bf r})~f(r_x,r_y) =
\lim_{\epsilon \rightarrow 0}
f({\epsilon \over \sqrt{2}},{\epsilon \over \sqrt{2}})
{}~.
\label{regb}
\ee

I am now ready to check that the quantum mechanical approach
discussed in this section can describe non-colliding ($w=0$) anyons.
Indeed the problem of two non-colliding anyons
in an harmonic potential
has been solved\cite{wil}, and in particular
it has been found that
\be
E_{2,0} = ( 10 + 2|\nu|)
{}~, \label{solb}
\ee
\be
|\u_{2,0}> =
N_{2,0} ~ r^{|\nu|}~
e^{ - {r^2 \over 2} + i l \phi }~
L_{2}^{|\nu|}(r^2)
{}~, \label{sola}
\ee
where $N_{2,0}$ is a normalization
constant.

\noindent
It is easy to verify that
Eqs.(\ref{solb}) and (\ref{sola}) are in perfect agreement
with Eqs.(\ref{eonecoppia}), (\ref{psioneto}), (\ref{etwocoppia}),
(\ref{etworega}), and (\ref{etworegb})
once the identifications (\ref{a-rel1})
are taken into account.

Note that, as shown in
Ref.[14],
the results obtained in this section for the two-body
Schr\"odinger problem can be easily generalized to $N$-body
Schr\"odinger problems; one finds that if $H_N$ is
the original Hamiltonian,
a renormalizable perturbation theory requires the use of the
Hamiltonian $H^{ren}_N$, given by
\be
H_N^{\delta} \equiv
H_{N} +
2 \pi g_b \sum_{m < n} \delta^{(2)}({\bf r_{n} - r_{m}}) ~.
\label{hndelgen}
\ee

Also note that my choice of
discussing eigenenergies and eigenfunctions
in this section instead of the
scattering amplitudes considered in the
preceding section was only an expedient to show
the general validity of these techniques. Indeed, as indicated by
results of
Ref.[12],
also the two-anyon s-wave scattering amplitude
can be calculated using this renormalized
perturbation theory in quantum mechanics.

\section{Conclusion}
I have discussed a field theoretical and a
quantum mechanical technique
for the investigation of colliding ($w \ne 0$)
and non-colliding ($w = 0$) anyons.

The results presented here
are also relevant to the issue
of which boundary conditions at the
points of overlap
are most natural in the case of anyons\cite{bour,mit2}.
Indeed, the results of Sec.2 and 3
establish descriptions of colliding anyons\cite{bour},
{\it i.e.} they identify the strength
of the contact coupling $g_r$ which (at a given scale)
is to be used in the calculations in order to describe anyons
whose wave functions satisfy the non-conventional
boundary conditions corresponding to
self-adjoint extension parameter $w$.
We have seen that
in the field theoretical approach (and
one can show that this holds also for
the quantum mechanical approach)
there appears to be no reason for restricting oneself
to the case of the conventional non-colliding
anyons\footnote{The
only special property of the
non-colliding anyons, which correspond to $w = 0$,
is the preservation of the scale invariance\cite{gacbak};
however, this property
is shared by the case of colliding anyons with $w \sim \infty$,
and, anyway,
there appears to be no physical motivation for
excluding values of $w$ that do not preserve scale invariance.},
the ones whose wave functions vanish
at the points of overlap.

As indicated by results presented in
Ref.[6],
the techniques here discussed for the investigation of anyons
can be rather straightforwardly generalized
to the case of ``non-abelian anyons" (also called
``NACS particles"\cite{gacbak,BJP}),
which are particles that can be described as bosons interacting
through the mediation of a non-abelian Chern-Simons gauge field.

\vglue 0.6cm
\leftline{\twelvebf Acknowledgements}
\vglue 0.3cm
I would like to thank Dongsu Bak, with whom I collaborated in
Ref.[6]
on which Sec.2 is based.
I acknowledge useful
discussions on the subject of this paper with Roman Jackiw
and Cristina Manuel.
This work is supported
in part by funds provided by the U.S. Department of Energy (D.O.E.)
under cooperative agreement \#DE-FC02-94ER40818,
and by Istituto Nazionale di
Fisica Nucleare (INFN, Frascati, Italy).

\vglue 0.6cm
\leftline{\twelvebf References}
\vglue 0.3cm

\end{document}